\documentclass[preprint,showpacs,superscriptaddress,preprintnumbers,amsmath,amssymb]{revtex4}

\usepackage{graphicx}
\usepackage{dcolumn}
\usepackage{bm}

\begin{document}


\title{Dynamics of Helping Behavior and Networks in a Small World}

\author{Hang-Hyun Jo} \email{kyauou2@kaist.ac.kr}
\affiliation{Department of Physics, Korea Advanced Institute of
Science and Technology, Deajeon 305-701, Republic of Korea}
\author{Woo-Sung Jung}
\affiliation{Department of Physics, Korea Advanced Institute of
Science and Technology, Deajeon 305-701, Republic of Korea}
\affiliation{Center for Polymer Studies and Department of Physics,
Boston University, Boston, MA 02215, USA}
\author{Hie-Tae Moon}
\affiliation{Department of Physics, Korea Advanced Institute of
Science and Technology, Deajeon 305-701, Republic of Korea}

\date{\today}

\begin{abstract}
To investigate an effect of social interaction on the bystanders'
intervention in emergency situations a rescue model was introduced
which includes the effects of the victim's acquaintance with
bystanders and those among bystanders from a network perspective.
This model reproduces the experimental result that the helping rate
(success rate in our model) tends to decrease although the number of
bystanders $k$ increases. And the interaction among homogeneous
bystanders results in the emergence of hubs in a helping network.
For more realistic consideration it is assumed that the agents are
located on a one-dimensional lattice (ring), then the randomness $p
\in [0,1]$ is introduced: the $kp$ random bystanders are randomly
chosen from a whole population and the $k-kp$ near bystanders are
chosen in the nearest order to the victim. We find that there
appears another peak of the network density in the vicinity of $k=9$
and $p=0.3$ due to the cooperative and competitive interaction
between the near and random bystanders.
\end{abstract}


\maketitle

\section{Introduction}

The concepts and methods of statistical physics and nonlinear
dynamics are applied to investigate the social, economic and
psychological phenomena
\cite{Weidlich1991,Mantegna2000,Vallacher1994}. Among the
interesting subjects that have attracted physicists are the opinion
dynamics \cite{Stauffer2005,Sznajd2000} including voting process
\cite{Alves2002,Bernardes2002} and social impact theory
\cite{Latane1981,Nowak1990,Lewenstein1992,Kohring1996,Plewczynski1998,Holyst2000}.
Social impact theory stemmed from the bystander effect by which
people are less likely to intervene in emergencies when others are
present than when they are alone as a result of the inhibitory
interaction among bystanders \cite{Latane1969,Amato1983}.

From the laboratory experiments about the emergency situations we
can gain an insight into this effect. When tested alone, subjects
behaved reasonably and the response rate was high. However the rate
was significantly depressed when they were with other subjects.
Subjects with others were unsure of what had happened or thought
other people would or could do something. In another experiment
subjects who were friends responded faster than those who were
strangers. The subjects who had met the victim were significantly
faster to report victim's distress than other subjects. And the
degree of arousal that bystanders perceive is assumed to be a
monotonic positive function of the perceived severity and clarity of
the emergency, and bystanders' emotional involvement with the victim
\cite{Piliavin1982}, which is also to be considered in an abstract
way in Section \ref{sect:2}.

In order to investigate the social phenomena as complex systems more
precisely we adopt the network point of view by which it means that
a social system consists of the interacting agents, where each node
and link of the network represent an agent and a relation or
interaction between a pair of agents respectively
\cite{Scott1991,Albert2002,Newman2003,Boccaletti2006}. A number of
properties about the real world networks such as social,
technological and biological ones, have been revealed and
investigated. Two of the main features of the real world networks
are the small world effect and the high clustering to be considered
in this paper by introducing a randomness $p$, the fraction of the
randomly chosen bystanders to the $k$ bystanders per accident, which
plays a similar role to that of Watts-Strogatz model
\cite{Watts1998}. The original dimensionless model for the bystander
effect is extended to the more realistic and general one in Section
\ref{sect:3}.

\section{\label{sect:2}Rescue Model}

Recently in order to investigate an effect of social interaction on
the bystanders' intervention in emergency situations a rescue model
(RM) was introduced \cite{Jo2006}. The model includes the effects of
the victim's acquaintance with bystanders and those among
bystanders. The RM focuses on the relations between agents rather
than on agents themselves, so defined is a relation spin between two
agents as $a_{ij}$ whose value is $1$ if agent $i$ (agent $j$) has
succeeded in rescuing agent $j$ (agent $i$), and $0$ otherwise.
$a_{ij}$ is symmetric and can be interpreted as an element of
adjacency matrix of helping network. Each agent $i$ has its
intervention threshold $c_i$ over which that agent can try to
intervene in an emergency situation.

At each time step an accident happens which consists of the degree
of the clarity or severity of accident represented as a random
number $q_v$ uniformly drawn from $[0,1]$, a randomly chosen victim
$v$ and $k$ bystanders which are also randomly chosen from a
population. $N_v$ denotes the set of $k$ bystanders. For each
bystander $i$ the degree of willingness to intervene $x_{vi}$ is
calculated:
\begin{equation}
x_{vi}(t)=q_v+\alpha a_{vi}(t) +\beta \sum_{j\in N_v, j\neq
i}{(2a_{ij}(t)-1)} -c_i .
\end{equation}
Only one bystander $i\in N_v$ with the largest value $x_{vi}$ can
intervene per accident, which can be called the intervener selection
rule. If we assume that the response speed of bystander $i$ is
exponential in $x_{vi}$, the selection of the bystander with the
largest $x_{vi}$ is justified. Additionally, once one bystander
intervenes, the pressures on the others will disappear. Then the
adjacency matrix is updated as following:
\begin{equation}
a_{vi}(t+1)=\theta \left(x_{vi}(t)\right)
\end{equation}
where $\theta(x)$ is a heaviside step function. If $x_{vi}\geq0$,
the rescue succeeds and then for the bystander $i$ who intervened,
the $a_{vi}$ gets the new value of one. In case of $x_{vi}<0$ the
rescue fails and then the $a_{vi}$ gets the new value of zero.
$\alpha$ represents the degree of victim's acquaintance with
bystander, so can be called an \textit{acquaintance strength}. The
third term of $x_{vi}$ is related to the interaction among
bystanders. $2a_{ij}-1$ gives $1$ if one bystander has succeeded in
rescuing the other or $-1$ otherwise. There does not exist any
neutral relation here. $\beta$ is used to tune the strength of
coupling so can be called a \textit{coupling strength}. Among them
the main control parameter is the number of bystanders $k$. As
observables we adopt the \textit{network density} \cite{Scott1991}
(helping rate in Ref. \cite{Jo2006}) and the \textit{success rate}
respectively:
\begin{equation}
a_k(t)=\frac{2}{N(N-1)}\sum_{i<j}{a_{ij}(t)},
\end{equation}
\begin{equation}
s_k=\frac{1}{T}\sum_{t=0}^{T-1}\theta (x_{vi}(t)). \label{eq:sk}
\end{equation}
In other words the success rate is defined as the number of
successful interventions divided by the total number of
interventions. Although the network density can be regarded as a
kind of helping rate, the success rate is closer to the helping rate
defined in the experimental studies \cite{Amato1983} in a sense that
the intervention may either succeed or fail without changing the
network density. We fix $c_i\equiv c=0.25$ for all $i$ according to
the experimental result \cite{Latane1969} that $70\sim 75\%$ of
isolated subjects intervened and $c$ does not change through this
paper, which means we consider a population composed of homogeneous
non-adaptive agents. Finally, the initial conditions are $a_{ij}=0$
for all pairs.

At first let us consider the case without the coupling effect among
bystanders, \textit{i.e.} $\beta =0$. Generally, an equation for the
network density can be written as \cite{Jo2006}
\begin{equation}
\frac{da_k(t)}{dt}=W_{0\rightarrow1}-W_{1\rightarrow0}, \label{eq:1}
\end{equation}
where
\begin{eqnarray*}
W_{0\rightarrow1}&=&(1-c)(1-a_k(t))^k, \\
W_{1\rightarrow0}&=&(c-\alpha)\left(1-(1-a_k(t))^k\right).
\end{eqnarray*}
$W_{0\rightarrow1}$ denotes the probability of creating a new link
between the victim and the bystander and $W_{1\rightarrow0}$ does
that of eliminating the existing link between them. The stationarity
condition for $a_k$ yields
\begin{equation}
a_k=1-\left(\frac{c-\alpha}{1-\alpha}\right)^{1/k}, \label{eq:ak}
\end{equation}
which says $a_k$ is a monotonically decreasing function of $k$. In
the numerical simulations $a_k(t)$ fluctuates around $a_k$ since the
links are added or removed with finite probabilities $1-c$ and
$c-\alpha$ respectively. As $k$ increases, so does the probability
that two connected agents, one as a victim and the other as a
bystander, get involved in an accident again. According to the
intervener selection rule one of the bystanders connected with the
victim must intervene and thus there is no reason for the increase
in $a_k$ according to $k$. Consequently the helping network gets
sparse with the number of bystanders.

An equivalent of the success rate defined in Eq. (\ref{eq:sk}) is
given by
\begin{eqnarray}
s_k&=&W_{0\rightarrow1}+W_{1\rightarrow1}\nonumber\\
&=&W_{0\rightarrow1}-W_{1\rightarrow0}+1-(1-a_k)^k=\frac{1-c}{1-\alpha},
\label{eq:2}
\end{eqnarray}
where we used the stationary solution for $a_k$ in Eq.
(\ref{eq:ak}). $s_k$ turns out to be independent of $k$ and of the
network density too. In fact, for the sparser network each link
should bear the more burden on the intervention to ensure the
success rate constant of $k$. From a viewpoint of the uncertainty of
a victim's receiving help from the bystanders $a_k$ corresponds to
the cost that the victim should pay to minimize the uncertainty.

\begin{figure}[!ht]
\centerline{\includegraphics[scale=1]{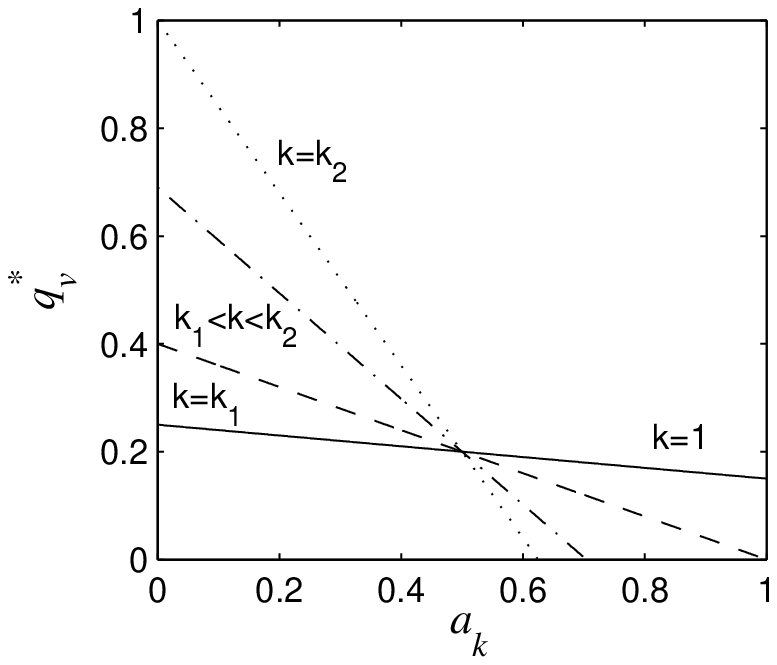}} \caption{The
diagram for the explanation of the existence of two transition
points $k_1$ and $k_2$, where $k_1=\frac{c-\alpha}{\beta}+1$ and
$k_2=\frac{1-c}{\beta}+1$ from the mean-field approximation Eq.
(\ref{eq:qcstar}). Here $\alpha=0.1$, $\beta=0.01$ and $c=0.25$ are
used.} \label{fig:mf}
\end{figure}

If the coupling effect among bystanders is taken into account, then
from the definition of $x_{vi}$ the condition for the successful
intervention can be obtained by a mean-field approximation,
\textit{i.e.} the substitution of $a_k$ for each $a_{ij}$:
\begin{equation}
x_{vi}=q_v+\alpha a_k +\beta(k-1)(2a_k-1)-c\ge0,\label{eq:xvi}
\end{equation}
or
\begin{equation}
q_v\ge -(\alpha+2\beta (k-1))a_k +\beta (k-1)+c\equiv
q_v^*.\label{eq:qcstar}
\end{equation}
At any time step, when given $a_k$ the success rate corresponds to
$1-q_v^*$. In case with $\beta > 0$ there appear two transition
points $k_1=\frac{c-\alpha}{\beta}+1$ and $k_2=\frac{1-c}{\beta}+1$
(see Fig. \ref{fig:mf}). At $k=k_1$ for any accident the rescue
succeeds, $s_k=1$, if and only if $a_k=1$ while at $k=k_2$ for any
accident the rescue fails, $s_k=0$, if and only if $a_k=0$. In the
range of $k_1\le k<k_2$, it is evident that $s_k\approx1$,
$q_v^*\approx0$ for $a_k\ge\frac{1}{2}
+\frac{c-\alpha/2}{\alpha+2\beta(k-1)}$. Once $c$ is larger than
$\alpha/2$, then the helping network is so dense that the
probability that the bystander who has not been connected with the
victim intervenes is extremely low, so is the possibility of
creating a new link. One can expect that $a_k(t)$ increases since
$s_k\approx1$, but very slowly since the network is sufficiently
dense.

Given $W_{1\rightarrow0}=0$ we can calculate the time evolution of
$a_k(t)$ by considering only the $W_{0\rightarrow 1}$. In case that
the victim is not connected with any of bystanders, if we assume
that at least one bystander is connected with all other bystanders,
then for $k_1\le k < k_2$,
\begin{eqnarray}
\frac{da_k(t)}{dt}&=&W_{0\rightarrow1}=(1-c-\beta(k-1))(1-a_k(t))^k\nonumber\\
&=&\beta(k_2-k)(1-a_k(t))^k.
\end{eqnarray}
Taking $a_k(t=0)=0$ as an initial condition yields
\begin{equation}
a_k(t)=1-\left[\beta(k_2-k)(k-1)t+1\right]^{-1/(k-1)}.
\label{eq:akt}
\end{equation}
Therefore $a_k(t\rightarrow\infty)=1$ for $k_1\le k < k_2$. This
solution represents the monotonically increasing behavior of the
network density with time step and the $k$ dependence as well.

\begin{figure}[!ht]
\centerline{\includegraphics[scale=1]{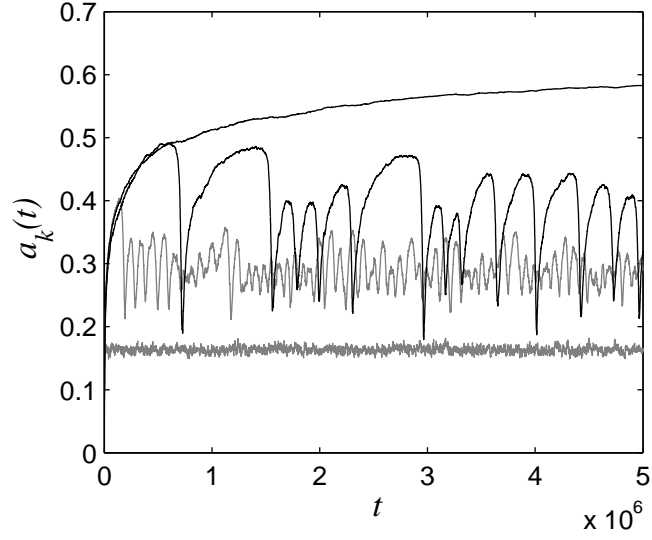}} \caption{The
numerical results of $a_k(t)$ for $k=9$ (lower gray line), $16$
(upper gray line), $17$ (lower black line) and $18$ (upper black
line) respectively. $a_{9}(t)$ fluctuates around some value.
$a_{16}(t)$ and $a_{17}(t)$ repeat the slow saturations punctuated
by the following abrupt declines. $a_{18}(t)$ increases
monotonically but very slowly and finally approaches $1$ as expected
in Eq. (\ref{eq:akt}). Here $N=100$, $\alpha=0.1$, $\beta=0.01$ and
$c=0.25$ are used.} \label{fig:akt}
\end{figure}

\begin{figure}[!ht]
\centerline{\includegraphics[scale=1]{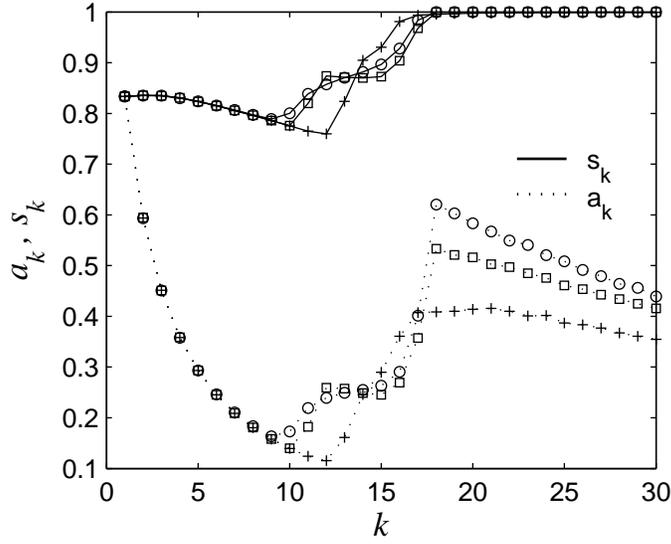}} \caption{The
numerical results of the network densities and success rates for
$N=100$ (circles), $N=500$ (squares) and $N=1000$ (plus signs)
respectively. Each data point of $s_k$ is obtained by averaging over
last $5\times 10^6$ time steps of entire $1.5\times10^7$ ($5\times
10^7$) time steps for $N=100$ (for $N=500, 1000$). And for $a_k$ the
time span is the same as $s_k$ but each point is averaged over every
$10^3$ time steps. For $k\geq k_1$ $a_k(t)$ increases monotonically
as expected in Eq. (\ref{eq:akt}), however the points are obtained
for finite time steps. Here $\alpha=0.1$, $\beta=0.01$ and $c=0.25$
are used.} \label{fig:ask}
\end{figure}

The time series of $a_k(t)$ shown in Fig. \ref{fig:akt} verify the
above arguments except that the transition occurs at $k=18$ larger
than $k_1$ expected by the mean-field approximation because of the
finite size effect. One can see from the Fig. \ref{fig:ask} that as
the system size increases, the transition point approaches $k_1=16$.
Additionally $a_k(t)$ exhibits the punctuated equilibrium-type
behaviors at $k$ slightly smaller than the transition point $k_1$,
which will be revised in relation to the network viewpoint.

Figure \ref{fig:ask} shows the numerical results for $s_k$ and
$a_k$, both of which decrease until $k$ reaches $9$ to $12$. This
tendency can be interpreted as the bystander effect in that the
bystanders are less likely to intervene in emergencies (succeed in
rescuing the victim in our model) when others are present than when
they are alone. Contrary to the case with $\beta=0$ the decreasing
$a_k$ according to $k$ has an additional negative effect on the
coupling among bystanders due to the positive $\beta$, thus lowers
the degrees of willingness $x_{vi}$ in Eq. (\ref{eq:xvi}) and
consequently $s_k$. However, $s_k$ and $a_k$ are getting large as
$k$ approaches $k_1$ because of the excitatory coupling among
bystanders.

Next, let us focus on the effects of the acquaintance strength
$\alpha$ and the coupling strength $\beta$ on the structure of
helping networks. If $\alpha=\beta=0$, since the degrees of
willingness for all bystanders are the same as $x_{vi}=q_v-c$, the
helping network shows a completely random structure. If we consider
the acquaintance effect, \textit{i.e.} $\alpha>0$, the probability
that two connected agents get involved in an accident again
increases. Therefore $\alpha$ has a `fixation' effect on the helping
network. If the coupling among bystanders is taken into account,
\textit{i.e.} $\beta>0$, the probability that the bystander
connected with more other bystanders is more likely to intervene in
an emergency, thus $\beta$ has an effect of `preferential attachment
(PA)' on the helping network. As a result of the PA the
heterogeneous hubs and hierarchical structures emerge from the
homogeneous non-adaptive population. The PA has been investigated
and summarized in Refs. \cite{Albert2002,Newman2003,Boccaletti2006}.

In addition, interestingly the above punctuated equilibrium-type
behaviors of $a_k(t)$ in Fig. \ref{fig:akt} accompany the rises and
falls of hubs when they undergo the slow saturations punctuated by
the abrupt declines. The nontrivial total collapse of helping
networks can result from the chain reaction between the effect of
cutting links due to the rescue failure and that of the rescue
failure due to the increasing negative interaction among bystanders.
This phenomenon is very different from those of other cases in which
once one agent becomes a hub, it lasts forever.

\section{\label{sect:3}Rescue Model in a Small World}

In the previous section we ignored the spatial property of the
system which does matter in realities. For more realistic
consideration the randomness $p \in [0,1]$ is introduced: when
assumed that the agents are located on a one-dimensional periodic
lattice (ring), the $k_r\equiv kp$ among $k$ bystanders are randomly
chosen from a whole population, which can be called the
\textit{random bystanders}, and the $k_n\equiv k-k_r$ bystanders are
chosen in the order in which they are nearest to the victim in the
Euclidean space, which can be called the \textit{near bystanders}.
The near bystanders are to the local neighborhoods what the random
ones are to the travelers from other places and so on. The
randomness $p$ makes the long-range interaction possible and plays
the similar role in our model to the randomness defined as a control
parameter of Watts-Strogatz small world networks \cite{Watts1998}.

\subsection{Agents in the One-dimensional World}

Let us first consider the case with $p=0$ which means that all the
bystanders are the near ones. In case of even $k$, one half of
bystanders are left to the victim and the other half are right to
the victim. In case of odd $k$, $k-1$ bystanders are chosen as for
the case of even $k$ except that the side of the last (farthest)
bystander is chosen randomly, that is, left or right to the victim.
We define a new observable $y_k$ as following:
\begin{equation}
y_k(t)=\frac{1}{\lceil k/2\rceil N}\sum_{i<j}{a_{ij}(t)}
\end{equation}
where $\lceil x \rceil$ is a ceiling function and the denominator is
the maximum number of links limited by the locality of interaction.
By the definition of $a_k$,
\begin{equation}
a_{k,p=0}=\frac{2}{N(N-1)}\sum_{i<j}a_{ij}= \frac{2\lceil
k/2\rceil}{N-1}y_k. \label{eq:p0p1}
\end{equation}
In case with $\beta=0$, since equations (\ref{eq:1})-(\ref{eq:2})
for the case without locality, \textit{i.e.} for $a_{k,p=1}$, are
valid for $y_k$, it is natural to regard $y_k$ as $a_{k,p=1}$. Thus
for small values of $k$ the network for the case limited by locality
becomes much sparser than that for the case without locality.
Interestingly $s_{k,p=0}$ turns out to be independent of $k$ again,
precisely $s_{k,p=0}=s_{k,p=1}=\frac{1-c}{1-\alpha}$. Similar to the
reason for the $k$ independence of $s_k$, in one-dimensional rescue
model the probability that two connected agents get involved in an
accident again is very high, thus the helping network gets sparse
and as a result each link bears more burden on the intervention.

\begin{figure}[!ht]
\centerline{\includegraphics[scale=1]{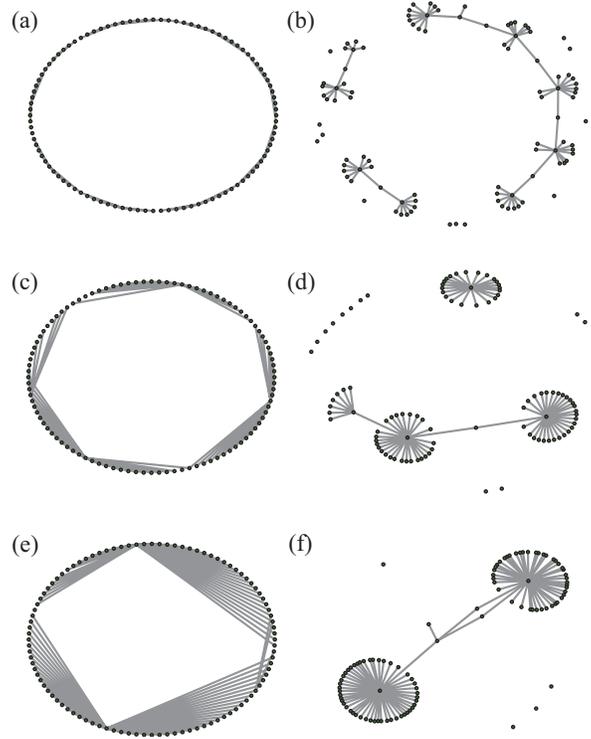}}
\caption{The numerical results of the helping networks in the
one-dimensional world for various $k$. The helping network for
$k=11$ is drawn in a circular style (a) and redrawn in (b). (c) and
(d) are for the case with $k=31$ and (e) and (f) are for the case
with $k=55$ respectively. The hubs emerge from the homogeneous
non-adaptive population. Here $N=100$, $\alpha=0.1$, $\beta=0.01$
and $c=0.25$ are used and the networks have been produced with the
Pajek software.} \label{fig:helpnet}
\end{figure}

In case with $\beta>0$ the helping networks in the one-dimensional
world consist of a few hubs induced by the PA effect and their
peripheries. The number of hubs amounts to about $N/k$ and the
number of peripheries per hub does to about $k$ as shown in Fig.
\ref{fig:helpnet}. Once the degrees of any agents become larger than
those of others by chance, they eventually grow to the hubs and
intervene in emergencies involved with their own peripheries and
vice versa, which forms some kind of helping communities. In
addition although the helping network in Fig. \ref{fig:helpnet} (b)
does not show the scale-free behavior of degree distribution its
backbone structure bears some resemblance to that of the structured
scale-free network \cite{Klemm2002} (see Fig. 8 in
\cite{Vazquez2003} for comparison).

\subsection{Agents in the Small world}

\begin{figure}[!ht]
\centerline{\includegraphics[scale=1]{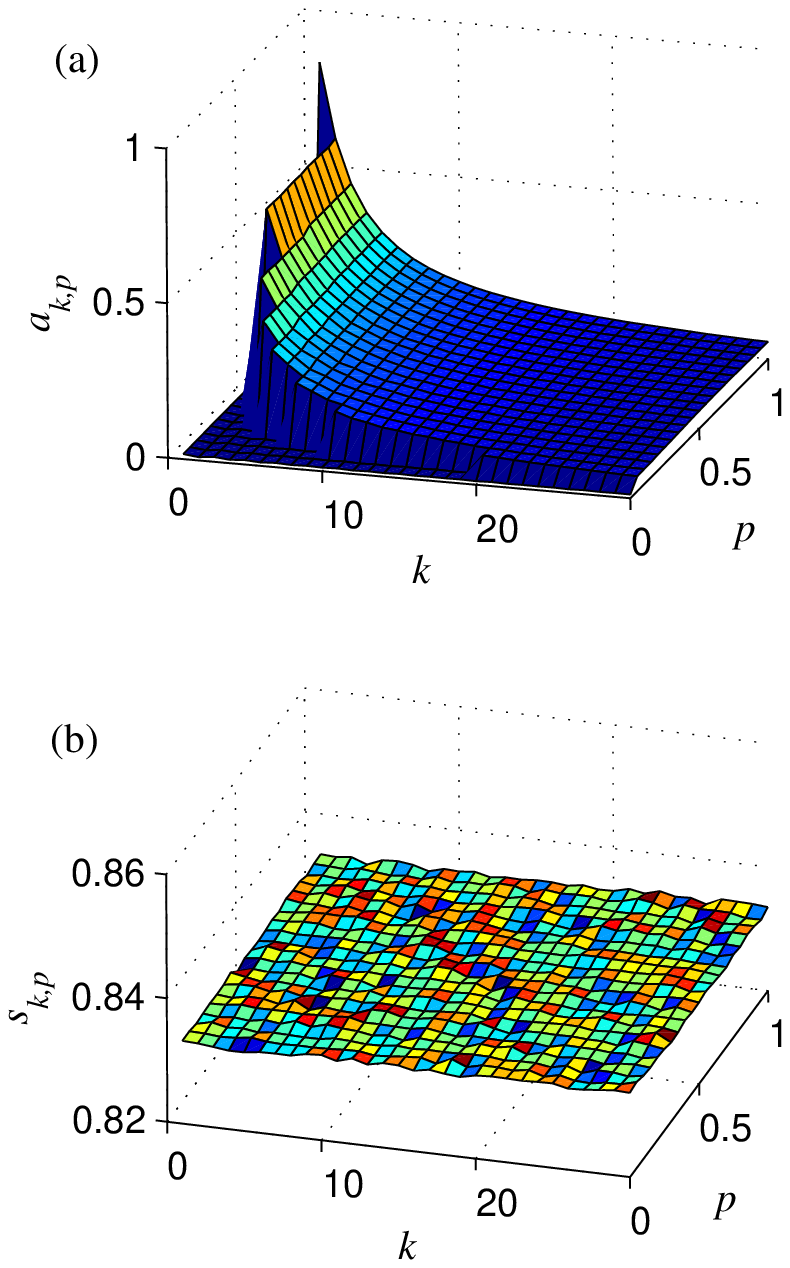}}
\caption{(Color online). The numerical results of the network
density and the success rate for $1\le k \le 30$ and $0\le p \le 1$
when $\beta=0$. Here $N=100$, $\alpha=0.1$ and $c=0.25$ are used.}
\label{fig:asb0}
\end{figure}

The network densities and the success rates are scanned for the
entire ranges of the number of bystanders $k$ and the randomness
$p$. When $\beta=0$ the numerical results depicted in Fig.
\ref{fig:asb0} show the trivial behaviors. For each $k$, according
to $p$ the network density $a_{k,p}$ leaps from $a_{k,p=0}$ to about
$a_{k,p=1}$ as soon as at least one random bystander appears, where
$kp_c=1$ or $p_c=1/k$. For the values of $p\ge p_c$ the network
densities rarely change regardless of $p$, which implies that what
is relevant is only whether the interaction is local or not and the
other factors do not matter. The uncertainty of receiving help is
maximized at $k=1$ and $p=1$, where there is only one bystander
chosen completely randomly per accident. Therefore the network
density should be maximized to ensure the success rate. As seen in
Fig. \ref{fig:asb0} (b), $s_{k,p}$ is independent of $k$ as well as
of $p$ since the coupling effect among bystanders is not taken into
account.

\begin{figure}[!ht]
\centerline{\includegraphics[scale=1]{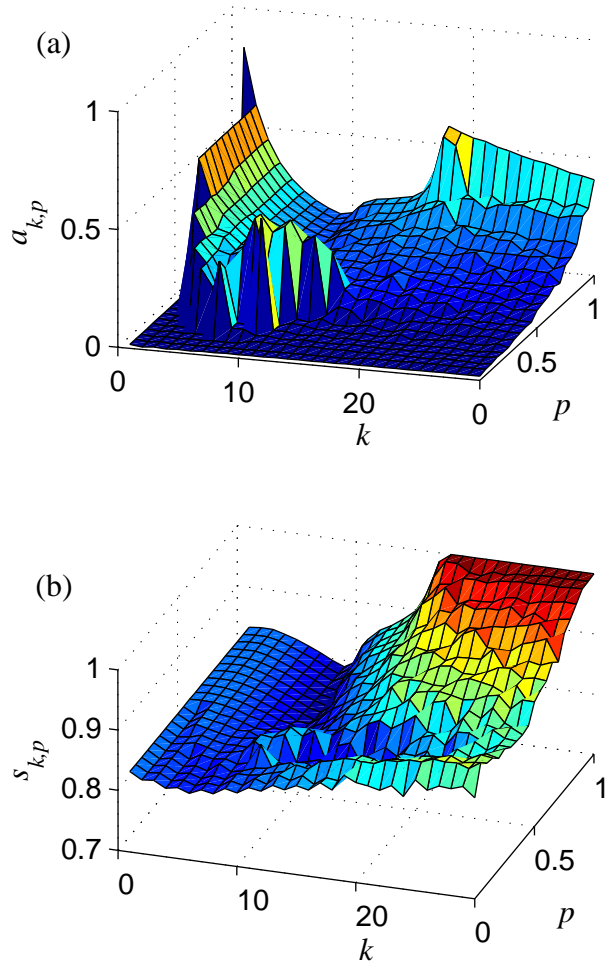}}
\caption{(Color online). The numerical results of the network
density and the success rate for $1\le k \le 30$ and $0\le p \le 1$
when $\beta=0.01$. It is found that there appears another peak of
$a_{k,p}$ and $s_{k,p}$ in the vicinity of $k=9$ and $p=0.3$. Here
$N=100$, $\alpha=0.1$ and $c=0.25$ are used.} \label{fig:asb001}
\end{figure}

If the coupling effect among bystanders is considered then an
interesting phenomenon is observed in Fig. \ref{fig:asb001}, that
is, there appears another peak of $a_{k,p}$ and $s_{k,p}$ in the
vicinity of $k=9$ and $p=0.3$. To understand the new peak in the
intermediate range of both $k$ and $p$ we focus on a cooperative or
competitive interaction between two groups; group of near bystanders
and that of random ones. Based on the relevance of positive $\beta$
we conjecture that the clustered structure of near bystanders is
essential to enhance the random bystanders' intervention and their
possibility of success so that the overall network density dominated
by the nonlocal links can grow to a large value. When $k$ fixed such
as $9$, for small $p$ (large $k_n$) local interactions among
clustered near bystanders dominate nonlocal interactions among
random ones and those across near and random ones, where the
locality restrains the network density from getting large. For
intermediate $p$ and $k_n$ the random bystanders are benefited from
the clustered near ones by making use of the existing links across
near ones and random ones near to the near ones and then the range
of interaction is expanded to the whole system after all.
Consequently the network density becomes large. Finally, for large
$p$ (small $k_n$) the near bystanders rarely cluster so that the
random ones cannot be benefited from the clustering of near ones,
hence the network density will converge to $a_{k,p=1}$ rather than
increase.

\begin{figure}[!ht]
\centerline{\includegraphics[scale=1]{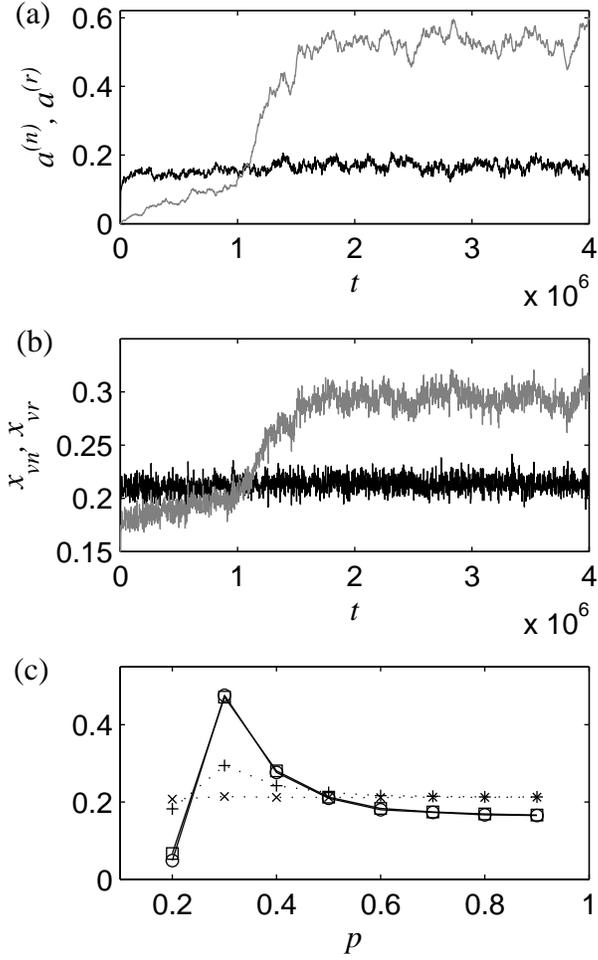}}
\caption{The numerical results for verifying the importance of
clustered near bystanders. (a) Time evolutions of two network
densities introduced in Eqs. (\ref{eq:newan})-(\ref{eq:newar});
$a^{(n)}(t)$ for near bystanders (black line) and $a^{(r)}(t)$ for
random ones (gray line). (b) Time evolutions of the degrees of
willingness; $x_{vn}(t)$ for near bystanders (black line) and
$x_{vr}(t)$ for random ones (gray line). For (a) and (b) each point
is averaged over $4000$ time steps. (c) The verification of Eq.
(\ref{eq:mfakp}) with $k=9$ by numerical simulations; $x_{vn}$
(crosses), $x_{vr}$ (plus signs), $a^{MF}_{k,p}$ (squares) and
$a_{k,p}$ (circles). Each point is averaged over $50$ realizations
after saturated. Since there is no random bystanders for $p\le 1/k$
and there is no near ones for $p=1$, in these ranges of $p$ either
$x_{vr}$ or $x_{vn}$ cannot be defined. Here $N=100$, $\alpha=0.1$,
$\beta=0.01$ and $c=0.25$ are used.} \label{fig:near_rand}
\end{figure}

To verify the above conjecture, at first two network densities are
newly introduced:
\begin{eqnarray}
a^{(n)}_{k,p}(t)&=&\frac{1}{Nk_n}\sum_{d(i,j)\leq
k_n}a_{ij}(t),\label{eq:newan} \\
a^{(r)}_{k,p}(t)&=&\frac{1}{N(N-1)/2-
Nk_n}\sum_{d(i,j)>k_n}a_{ij}(t),\label{eq:newar}
\end{eqnarray}
where $d(i,j)$ gives the shorter distance on the ring between agent
$i$ and $j$ and the superscripts $n$ and $r$ represent the near and
random bystanders respectively. $Nk_n$ in Eqs.
(\ref{eq:newan})-(\ref{eq:newar}) is the maximum number of links
among near bystanders. From the extensive numerical simulations it
is found that for the values of $k$ and $p$ other than the new peak
region and the growth region of $k_1 \leq k< k_2$ and $p$ near to
$1$, both $a^{(n)}(t)$ and $a^{(r)}(t)$ fluctuate around some
values. On the other hand, for the new peak region, precisely at
$k=9$ and $p=0.3$, in Fig. \ref{fig:near_rand} (a) $a^{(n)}(t)$
jumps to about $0.15$ very quickly and fluctuates around that value
for a while until $a^{(r)}(t)$ grows exponentially to exceed
$a^{(n)}(t)$, which is called the intersection point. $a^{(r)}(t)$
increases fast then saturates while $a^{(n)}(t)$ also increases a
little. To figure out whether two bystander groups cooperate or
compete we calculate the cross-correlations between two network
densities before and after the intersection point. Before that point
the cross-correlation is $0.5236$ while after that point it is
$-0.3565$. In the early stage the near bystanders help the random
ones intervene but once the random ones dominate the near ones, two
groups compete for the chance of intervention.

For more specific investigation we calculate the averaged degree of
willingness of random bystanders $x_{vr}(t)$ as a function of time
and that of near ones $x_{vn}(t)$ respectively as shown in Fig.
\ref{fig:near_rand} (b). Similar to the time evolution of
$a^{(r)}(t)$, $x_{vr}(t)$ increases from a lower value than
$x_{vn}(t)$ then exceeds it at the intersection point while
$x_{vn}(t)$ fluctuates around some value. $x_{vr}$ larger than
$x_{vn}$ implies the more chance of random bystanders' intervention
than that of near ones' intervention. As a result the success rate
of random ones ($0.8531\pm 0.0101$) is also slightly larger than
that of near ones ($0.8212\pm 0.0221$). The difference of two
averaged degrees of willingness affects which kind of bystanders are
more likely to intervene and succeed, which affects the network
density and the success rate successively. For general $k$ by
mean-field approximation the degrees of willingness are given by
\begin{eqnarray*}
x_{vr}&=&q_v+\alpha a_{k,p}+\beta (k-1)(2a_{k,p}-1)-c, \\
x_{vn}&=&q_v+\alpha a_{k,p=1}+\beta (k-1)(2a_{k,p=1}-1)-c,
\end{eqnarray*}
where we have assumed that the probability that a random bystander
is connected with any other bystander is the same as $a_{k,p}$ and
for the near bystanders the probability to be connected with any
other one is $y_k=a_{k,p=1}$. From these the mean-field network
density is obtained:
\begin{eqnarray}
a^{MF}_{k,p}=a_{k,p=1}+\frac{x_{vr}-x_{vn}}{\alpha+2\beta (k-1)}.
\label{eq:mfakp}
\end{eqnarray}
The numerical results in Fig. \ref{fig:near_rand} (c) for Eq.
(\ref{eq:mfakp}) with $k=9$ indicate that the mean-field approach
works.

For the other values of $k$ what happens according to the randomness
$p$? For smaller values of $k$ and $p\geq p_c$ the clustering of
near bystanders rarely contributes to the random ones' intervention
because the network is relatively dense to ensure the success rate
for small $k$. Therefore it is the locality that is relevant for the
results as for the case with $\beta=0$. For larger values of $k$,
especially larger than $k_1$, we already know the network density
goes to $1$ but very slowly when $p=1$. Since for small $p$ the
network density converges to some value, there should be a
transition line where the clustering cannot exist, \textit{i.e.}
$k_n\leq 1$, equivalently $p\geq 1-1/k\equiv p^*$. When $p<p^*$ and
$k\geq k_1$ the locality limits the long-range excitatory
interaction among bystanders and at the same time the probability
that the victim and the bystander are already connected is
relatively high due to large $k$ so that it is difficult for near
bystanders to form an effective cluster and thus to enhance the
random ones' intervention too.

Next, as initial conditions we take the random networks whose
network density $a(0)$ varies from $0.05$ to $0.5$. Only for the
small value of $a(0)$, precisely $0.05$, there appears the tiny peak
in the intermediate range of $k$ and $p$ while no peak appears in
the other cases with larger $a(0)$. This is because the initial
random structure does not allow any structural change for the
clustering of near bystanders. In conclusion, the new peak in an
intermediate range of $k$ and $p$ has been made sense by introducing
the clustering effect of near bystanders.

\begin{figure}[!ht]
\centerline{\includegraphics[scale=1]{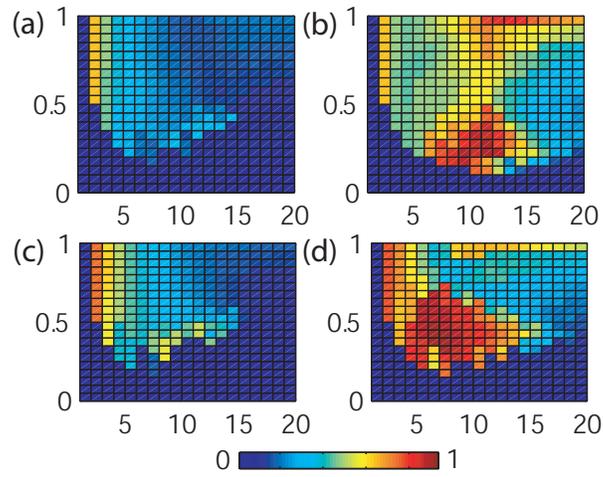}}
\caption{(Color online). The numerical results of the landscape of
network density for $1\leq k \leq 20$ (horizonal axis) and $0\leq
p\leq 1$ (vertical axis), changing from the sharp peak to a plateau
according to $\alpha$ and $\beta$: (a) $\alpha=0.05$, $\beta=0.005$,
(b) $\alpha=0.05$, $\beta=0.02$, (c) $\alpha=0.2$, $\beta=0.005$,
and (d) $\alpha=0.2$, $\beta=0.02$. Each point is averaged over $10$
realizations. Here $N=100$ and $c=0.25$ are used.}
\label{fig:alphabeta}
\end{figure}

Finally it is also observed in Fig. \ref{fig:alphabeta} that
increasing $\alpha$ and $\beta$ lead to the overall increase in the
height of peak of network density then change the shape of peak from
a sharp one to a plateau. Since $\alpha$ has a fixation effect on
the existing victim-bystander link it cannot contribute to the
creation of new link but only can lower the possibility of deleting
the existing link. On the other hand $\beta$ in relation to the
bystander-bystander links can lead to the creation of new link
between victim and bystander and the deletion of existing link
depending on the network density. Therefore around the peak
increasing $\beta$ enhances the successful interventions of random
bystanders based on the clustered near ones, then raises the network
densities, which are also preserved by increasing $\alpha$.


\section{Conclusions}
In this paper we have studied not only the original rescue model,
which was introduced in order to investigate an effect of social
interaction on the bystanders' intervention in emergency situations,
but also the rescue model on a small world. The bystander effect has
been successfully reproduced from numerical simulations and
explained by the mean-field approximation. In general both of the
local interaction and the increasing $k$ reduce the network density
since the victim has more chance to get involved in the acquainted
bystander. However, it is found that when the coupling effect among
bystanders considered there appears another peak of $a_{k,p}$ and
$s_{k,p}$ in the vicinity of $k=9$ and $p=0.3$ for some given
parameters, which results from the enhancement of nonlocal
interventions based on the clustering effect of near bystanders.

The relation spins $a_{ij}$ compose the helping networks. The
coupling effect represented by positive $\beta$ induces the
emergence of hubs from a homogeneous non-adaptive population. In the
original rescue model the rises and falls of hubs have been observed
near the transition point $k_1$ and in one-dimensional world the
whole population is divided into a few helping communities, each of
which consists of a hub and its peripheries. Although we could not
find any real world helping networks to our knowledge, these results
give us an insight into the dynamics of helping behavior and
networks.

\begin{acknowledgments}
The authors thank Jae-Suk Yang, Eun Jung Kim, and Pan-Jun Kim for
fruitful discussions.
\end{acknowledgments}



\newpage


\end{document}